\documentclass[12pt]{iopart}

\usepackage{graphicx}
\usepackage{subfig}
\usepackage{amssymb}
\usepackage{amsthm} 
\usepackage{mathptmx}
\usepackage{nicefrac}
\usepackage{xcolor}
\begin{document}

\title[]{A new kaonic helium measurement in gas by SIDDHARTINO at the DA$\Phi$NE collider}

\author{ D Sirghi$^1$, F Sirghi$^1$, F Sgaramella$^{1*}$,  M Bazzi$^1$, D Bosnar$^2$, M Bragadireanu$^3$, M Carminati$^4$, M Cargnelli$^5$, A Clozza$^1$, G Deda$^4$, L De Paolis$^1$, R Del Grande$^{1,6}$, L Fabbietti$^6$, C Fiorini$^4$, C Guaraldo$^1$, M Iliescu$^1$, M Iwasaki$^7$, P Levi Sandri$^1$, J Marton$^5$, M Miliucci$^1$, P Moskal$^8$, F Napolitano$^1$, S Nied\'{z}wiecki$^8$, K Piscicchia$^{9,1}$, A Scordo$^{1**}$, H Shi$^5$, M Skurzok$^8$, M Silarski$^8$, A Spallone$^1$, M T\"uchler$^5$, O Vazquez Doce$^1$, J Zmeskal$^5$ and C Curceanu$^1$}

\address{$^1$ Laboratori Nazionali di Frascati INFN, Via E. Fermi 54, 00044 Frascati, Italy}
\address{$^2$ Department of Physics, Faculty of Science, University of Zagreb, Zagreb, Croatia}
\address{$^3$ Horia Hulubei National Institute of Physics and Nuclear Engineering (IFIN-HH) Măgurele, Romania}
\address{$^4$ Politecnico di Milano, Dipartimento di Elettronica, Informazione e Bioingegneria and INFN Sezione di Milano, Milano, Italy}
\address{$^5$ Stefan-Meyer-Institut f\"ur Subatomare Physik, Vienna, Austria}
\address{$^6$ Excellence Cluster Universe, Technische Universiät München Garching, Germany}
\address{$^7$ RIKEN, Tokyo, Japan}
\address{$^8$ Faculty of Physics, Astronomy, and Applied Computer Science, Jagiellonian University, {\L}ojasiewicza~11, 30-348 Krak\'{o}w, Poland}
\address{$^9$ Centro Ricerche Enrico Fermi – Museo Storico della Fisica e Centro Studi e Ricerche “Enrico Fermi”, Via Panisperna 89A 00184, Roma, Italy}

\ead{$^*$ francesco.sgaramella@lnf.infn.it (Corresponding Author)}
\ead{$^{**}$ alessandro.scordo@lnf.infn.it (Corresponding Author)}

\vspace{10pt}
\begin{indented}
\item[] January 2022
\end{indented}

\begin{abstract}

\noindent The SIDDHARTINO experiment at the DA$\Phi$NE Collider of INFN-LNF, the pilot run for the SIDDHARTA-2 experiment which aims to perform the measurement of kaonic deuterium transitions to the fundamental level, has successfully been concluded. The paper reports the main results of this run, including the optimization of various components of the apparatus, among which the degrader needed to maximize the fraction of kaons stopped inside the target, through measurements of kaonic helium transitions to the 2p level. 
The obtained shift and width values are $\mathrm{\epsilon_{2p}= E_{exp} - E_{e.m} = 0.2\pm2.5 (stat)\pm2 (syst)\ eV }$ and $\mathrm{\Gamma_{2p} = 8\pm10\ eV \,(stat)}$, respectively. This new measurement of the shift, in particular, represents the most precise one for a gaseous target and is expected to contribute to a better understanding of the kaon-nuclei interaction at low energy. \\
\newline
\emph{We dedicate this article to the memory of a colleague and friend George Beer, with whom some of us shared the adventure of strangeness physics for many years.}
\end{abstract}

%
\vspace{2pc}

\section*{Introduction}

Light kaonic atoms spectroscopy is a unique tool for the investigation of the low-energy strangeness Quantum Chromodynamics (QCD). Precise measurements of the radiative X-ray transitions towards low-n levels of these systems provide information on the kaon-nucleus interaction at threshold which, in typical scattering experiments, would require an extrapolation towards zero energy, making them method-dependent.\\
In this context, a special role is played by the lightest kaonic atoms, namely kaonic hydrogen, deuterium, and helium. From the first two, the isospin-dependent antikaon-nucleon scattering lengths can be obtained from the measurements of the strong interaction induced shifts and widths of the 1s levels. Additional information on the strong interaction with many-body systems can be retrieved from transitions to the 2p level of kaonic helium 3 and 4 \cite{revmodphys:2019}.\\
The kaonic hydrogen 1s level strong interaction induced shift and width have been successfully measured by the SIDDHARTA experiment in 2009 \cite{Bazzi:2011}.\\
For what regards the effect of the strong interaction on the 2p level of kaonic helium isotopes, the pool of experimental data is rich. The E570 experiment at KEK, in Japan, delivered in 2007 the first measurement of the $K^4He(3d\rightarrow2p)$ transition in liquid helium \cite{Okada:2007ky}, which was followed by the first measurement with a gaseous target obtained by SIDDHARTA together with the first measurement ever of $K^3He(3d\rightarrow2p)$ \cite{SIDDHARTA:2012rsv}.\\ 
The kaonic deuterium measurement is a very challenging one, due to the fact that the yield of the X-ray transitions to the 1s level is expected to be about one order of magnitude lower than that of kaonic hydrogen. This measurement will be performed for the first time by the SIDDHARTA-2 experiment, presently under installation at the DA$\Phi$NE collider \cite{Zobov:2010zza,Milardi:2018sih}, scheduled to take data in 2022-2023.
Before the SIDDHARTA-2 kaonic deuterium measurement, a pilot run with a reduced setup, SIDDHARTINO, was performed in 2021 during the commissioning of the DA$\Phi$NE collider.\\
The aim of this run was twofold: on one side, the optimization of the run conditions, including the collider luminosity measurement, the trigger system, and the X-ray Silicon Drift Detectors (SDDs) \cite{Miliucci:2021vil}; on the other side, the optimization of the degrader used by the experiment to maximize the fraction of kaons stopped inside the gaseous helium target.\\
This second step was performed by measuring kaonic helium-4 transitions on the 2p level, which has a much higher yield \cite{Bazzi:2014} than the kaonic deuterium expected one.\\
The pilot SIDDHARTINO run also allowed to obtain the most precise measurement of kaonic helium transitions in a gaseous target, which is reported in this paper together with all the optimizations which allowed to obtain this record.\\
In Section 1, the experimental apparatus is described, while in Section 2 we report the detector calibration procedure, the background suppression tools, and the kaon degrader optimization. In Section 3, the results for the shift and the width of the kaonic helium-4 2p level are presented, while in Section 4 the conclusions can be found.

\section{SIDDHARTINO at the DA$\Phi$NE collider }\label{setup_section}

\begin{figure}[htbp]
\centering
\mbox{\includegraphics[width=10 cm]{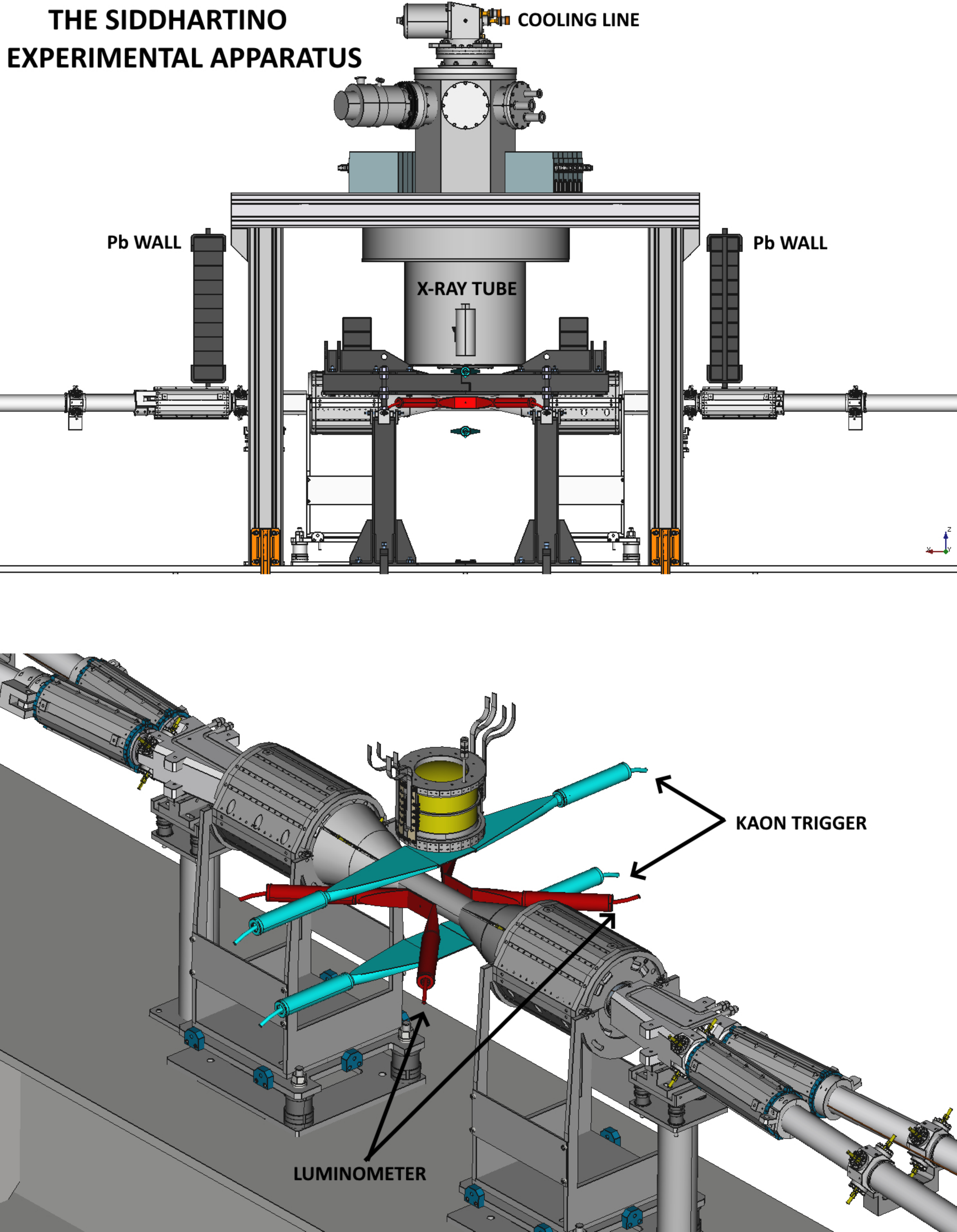}}
\caption{Top: schematic view of the SIDDHARTINO experimental apparatus installed on the IP of the DA$\Phi$NE collider. Bottom: view of the target cell, the kaon trigger (cyan), and the luminometer (red) systems.}
\label{siddhino_new}
\end{figure}

DA$\Phi$NE \cite{Zobov:2010zza,Milardi:2018sih} is a world-class double ring $e^+e^-$ collider operating at the center of mass energy of the $\phi$-resonance ($\mathrm{m_{\phi}= 1.02\, GeV/c^2}$). The back-to-back $K^+K^-$ pairs, resulting from its almost at-rest decay with a $\simeq 50\%$ branching ratio \cite{Zyla:2020zbs}, are characterized by low momentum ($\simeq$ 127 MeV/c) and low energy spread ($\delta p/p\simeq 0.1\%$), making this machine an ideal environment to perform precision spectroscopy measurements of kaonic atoms.\\
The SIDDHARTA-2 experiment at DA$\Phi$NE aims to perform the first measurement ever of the X-ray transitions to the fundamental level of kaonic deuterium which, combined with the kaonic hydrogen one performed by SIDDHARTA \cite{Bazzi:2011}, will be used to obtain the antikaon-nucleon isospin dependent scattering lengths \cite{Doring:2011,Sheva:2017}, fundamental quantities for a better understanding of the strong interaction theory in the strangeness sector. Before performing this challenging measurement, during the DA$\Phi$NE collider commissioning phase in 2021, a pilot run was performed with a reduced version of the setup, SIDDHARTINO, installed in the interaction point (IP) of the accelerator. The goal of this run was to assess and optimize the performances of the machine and the experimental apparatus via a measurement of the kaonic helium transitions to the 2p level, made possible by the much higher yields of these transitions compared to the kaonic deuterium ones.\\
A sketch of the SIDDHARTINO experimental apparatus is shown in Fig. \ref{siddhino_new}. 

\noindent A cylindrical vacuum chamber evacuated below $\mathrm{10^{-5}\, mbar\,}$, is placed above the DA$\Phi$NE IP and contains the target cell. The closed-cycle helium refrigerator cooling system, shown in the upper part of Fig. \ref{siddhino_new}, is used to cool the target cell down to $\mathrm{25\, K}$. A system composed of two X-ray tubes is employed for the in-situ calibration of the Silicon Drift Detectors (SDDs); these are used to measure the X-rays emitted from the radiative transitions of kaonic atoms. The calibration takes advantage of the excitation of the fluorescence lines of high purity Ti-Cu targets. Finally, the two lead walls in Fig. \ref{siddhino_new} represent the shielding structures used to shelter the apparatus from the particles,  mostly Minimum Ionizing Particles (MIPs), lost from the $e^+e^-$ rings mainly by the Touschek effect.\\
In the lower part of Fig. \ref{siddhino_new} the kaon trigger (KT) and the luminometer systems, both based on pairs of scintillators read at both ends by photomultipliers (PMs), as well as the target cell, are shown. 
The luminometer, realized by the SIDDHARTA-2 collaborators of the Jagiellonian University (red) is employed to evaluate both the machine luminosity and background via time of flight measurements of kaons and MIPs on the horizontal plane \cite{Skurzok:2020phi}. The $K^+K^-$ pairs emitted from the $\phi$ decay on the vertical plane are detected by the KT (cyan) before entering the target cell. The top scintillator is placed just in front of the entrance window of the vacuum chamber and it acts as an efficient asynchronous background rejection tool providing a narrow timing window to tag X-ray events in coincidence with kaons reaching the target. \\
The target cell is a cylinder of 144 mm in diameter and 125 mm in height, made of high purity aluminum bars and 150 $\mathrm{\mu m}$ thick Kapton walls. It is placed inside the vacuum chamber above the beam pipe, it has a 125 $\mathrm{\mu m}$ entrance window, a 100 $\mathrm{\mu m}$ thick Ti top window and it is surrounded by 8 SDD arrays, with a dedicated thermal contact to keep them stable at the $\mathrm{170\, K}$ working temperature.\\
The large area monolithic SDD arrays have been developed by Fondazione Bruno Kessler (FBK, Trento), Politecnico di Milano (PoliMi), Laboratori Nazionali di Frascati (INFN-LNF), and the Stefan Meyer Institute (SMI, Vienna) specifically to be employed by the SIDDHARTA-2 experiment to perform high-precision spectroscopy of light kaonic atoms. Each 450 $\mathrm{\mu m}$ thick silicon array consists of 8 single cells arranged in a 2 × 4 matrix for a total active area of 5.12 $\mathrm{cm^2}$. The silicon wafer is glued on an alumina carrier which provides the polarization to the units via an external voltage. The charges generated by the X-ray absorption within the silicon bulk are collected by a point-like central anode and amplified through a closely bonded CMOS low-noise, charge-sensitive preamplifier. The signals are then processed by a dedicated ASIC which filters them through a 2 $\mathrm{\mu s}$, 9th order semi-Gaussian shaping stage to minimize the electronic noise contribution  \cite{Quaglia:2016uox,Schembari:2016IEE}. Each ASIC handles the signals produced by 16 units and provides the data acquisition (DAQ) chain with the individual amplitude and timing information. The spectroscopic performances of this system have been optimized \cite{Miliucci:2019mdpi} and tested in the first DA$\Phi$NE commissioning phase where, with an energy resolution of $\mathrm{157.8\pm0.3^ {+0.2}_{-0.2}\,eV}$ at $\mathrm{6.4\,keV}$ and a linearity at the level of 2-3 eV \cite{Miliucci:2021wbj}, they proved to be suitable to perform high precision kaonic atoms measurements.\\
The second commissioning phase has been devoted to the fine tuning of the kaon degrader system (see Section \ref{deg_section}), one of the crucial components of the experiment. The purpose of the degrader is to slow down the kaons before entering the target cell to optimize their stopping distribution inside the gas target. For the optimization of the degrader, the SIDDHARTINO target cell has been filled with $^4He$ gas at a temperature of $\mathrm{25\, K}$ and pressure of $\mathrm{1\, bar}$, which corresponds to $1.5\,\%$ of liquid helium density (LHeD). The high yield of the $3d\rightarrow2p$ transition of kaonic helium-4 ($K^4He$) allowed to appreciate variations in the number of signal events at the level of $\simeq\,20\,\%$ in a few hours of data taking. The final period of the SIDDHARTINO data taking in June 2021 was dedicated to the measurement of $K^4He(3d\rightarrow2p)$ transitions at lower density, namely 0.4 bar, corresponding to 0.73 $\%$ LHeD.
In this work we present the results of the analysis performed on the summed spectra of 16.5 $\mathrm{pb^{-1}}$ and 9.5 $\mathrm{\,pb^{-1}}$ for the 1.5$\%$ and 0.73$\%$ LHeD runs, respectively, for a total integrated luminosity of 26 $\mathrm{\,pb^{-1}}$.

\section{Data analysis and degrader optimization}

\subsection{Calibration procedure}\label{calib_section}
\begin{figure}[htbp]
\centering
\mbox{\includegraphics[width=13 cm]{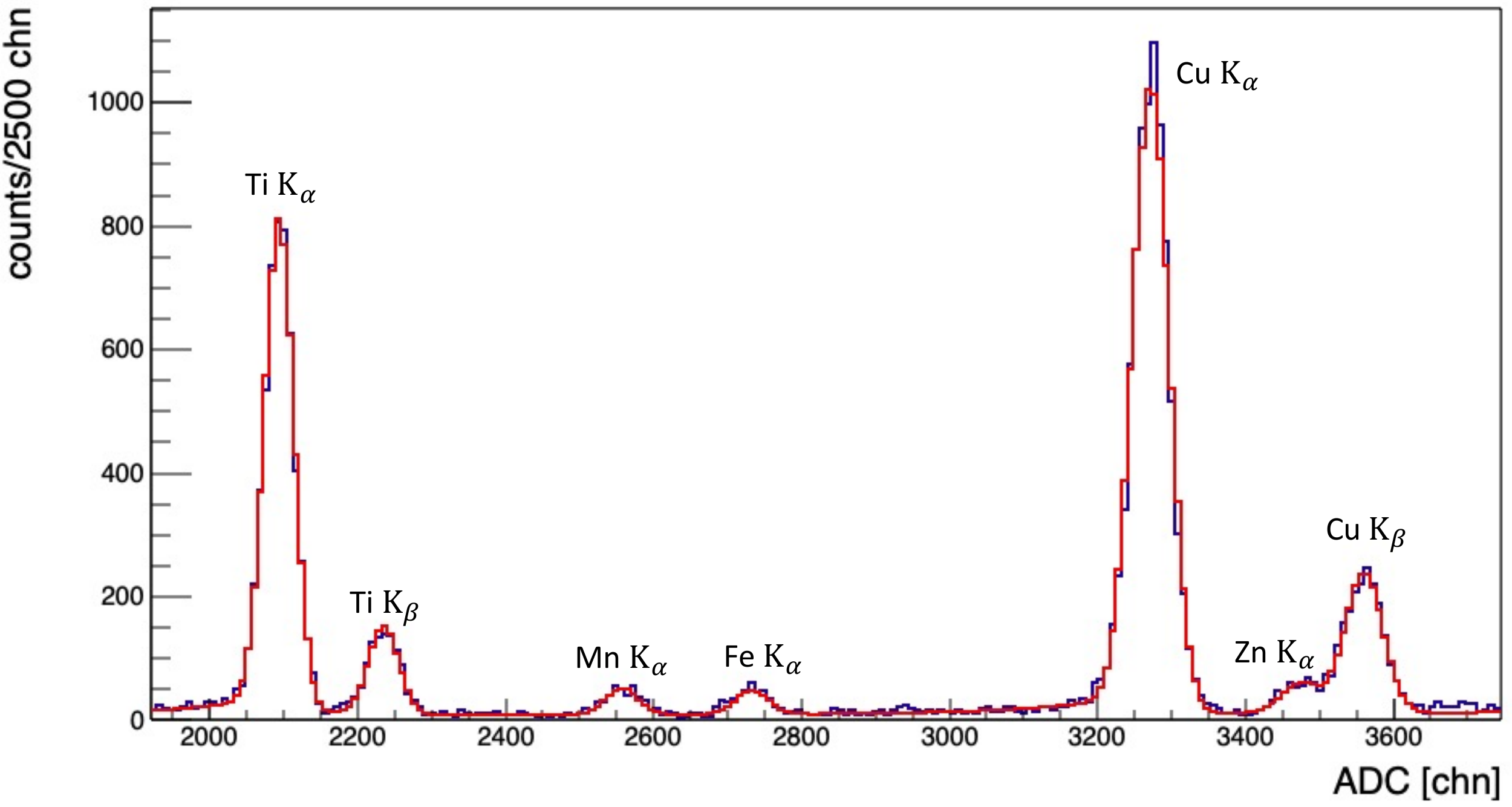}}
\caption{\small{Typical spectrum for a single SDD obtained after a calibration run ($\sim$ 1 hour of data taking); the calibration peaks (Ti and Cu) were fitted using a Gaussian function plus an exponential low energy tail. Mn, Fe and Zn come from the accidental excitation of various components of the experimental apparatus. The unit scale is given by the Analog to Digital Converter (ADC).}}
\label{calib}
\end{figure}

\noindent The energy calibration of the SDDs is one of the most critical aspects of the whole data analysis procedure. It was performed using characteristic emission lines induced by the system of two X-ray tubes on high purity titanium and copper strips placed on the target cell walls.
Since SDD cells and their associated front-end electronics are characterized by different charge collection and voltage conversion functions, individual calibrations are mandatory before summing up all the measured spectra \cite{Miliucci:2021wbj}. Periodic calibration runs during the whole data taking period were thus performed to obtain the energy response of the single detectors.
Fig. \ref{calib} shows a typical spectrum for a single SDD obtained in a calibration run ($\sim$ 1 hour). The peaks present in the spectrum, corresponding to the X-rays emitted by the activated titanium and copper strips, were fitted using a Gaussian plus an exponential low energy tail to reproduce the SDD response function \cite{Gysel:2003}. The contribution of the low energy tail to the peak amplitude was found to be less than 1\% \cite{2021tail}. Together with the calibration lines, other peaks are also visible (Mn, Fe, and Zn), resulting from the accidental excitation of various components of the experimental apparatus, which have been fitted with the same functions.

\subsection{Background suppression}\label{bkg_section}

\begin{figure}[htbp]
\centering
\includegraphics[width=12.0 cm]{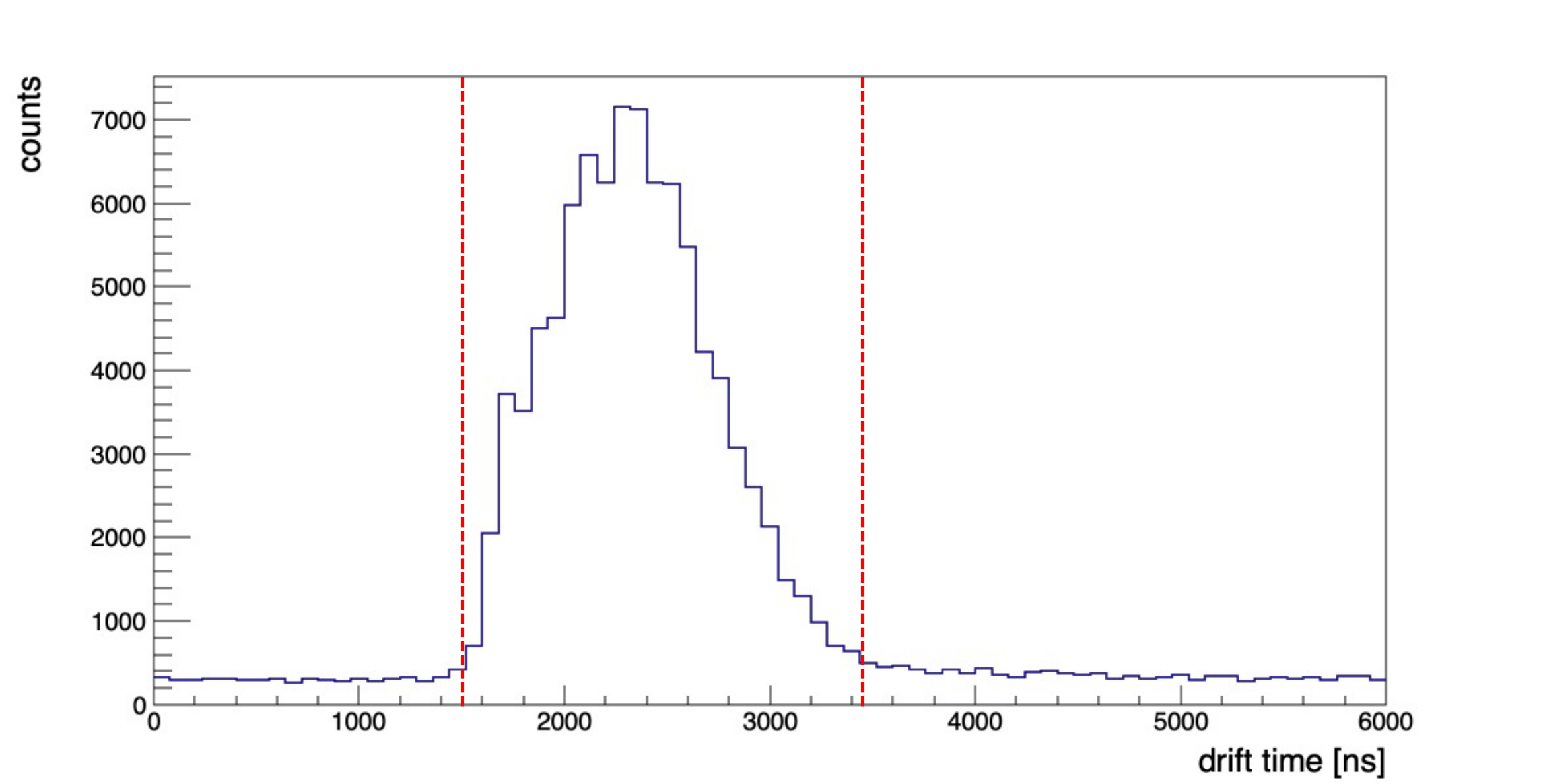} 
\includegraphics[width=12.0 cm]{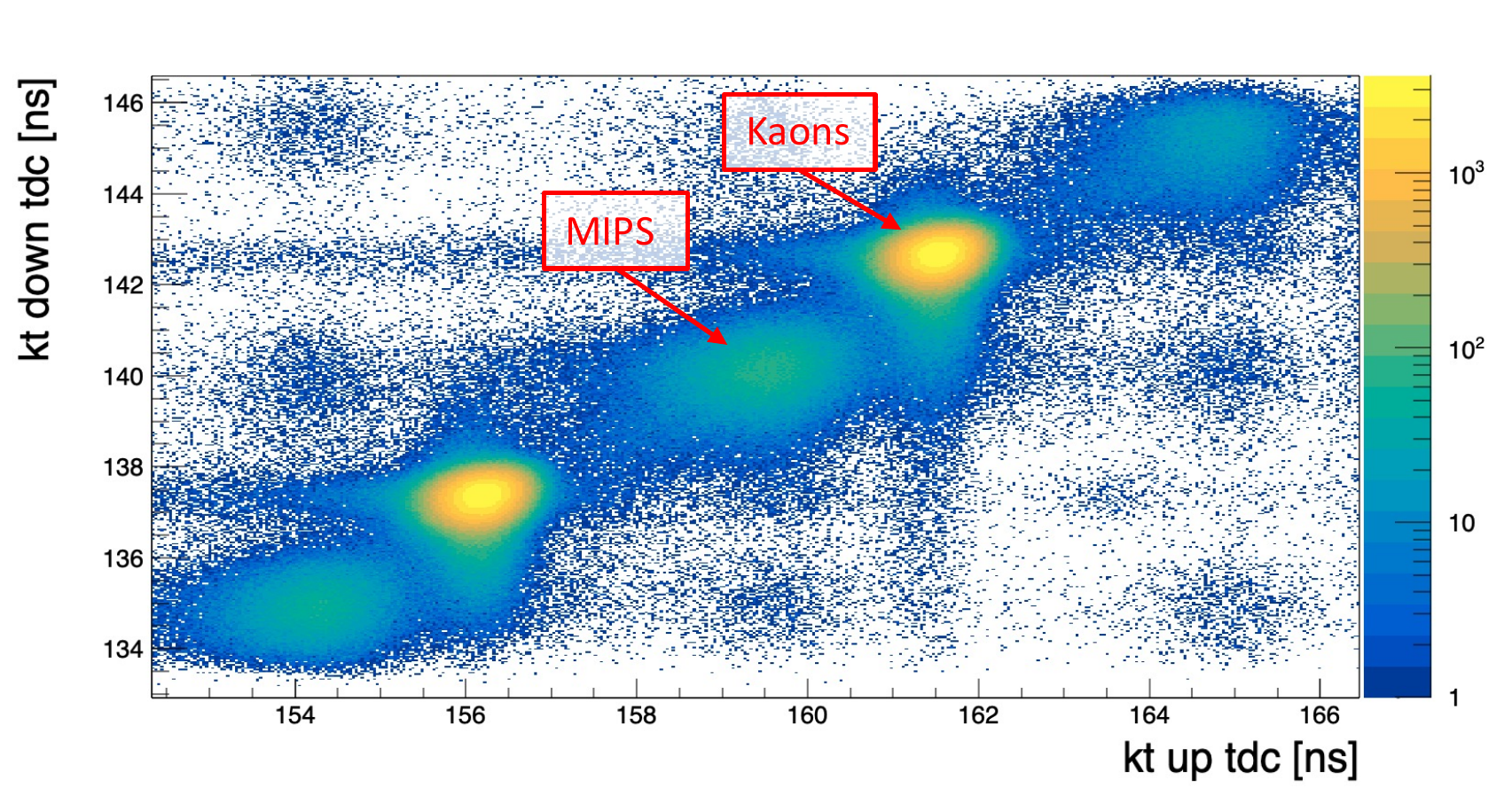} 
\includegraphics[width=12.0 cm]{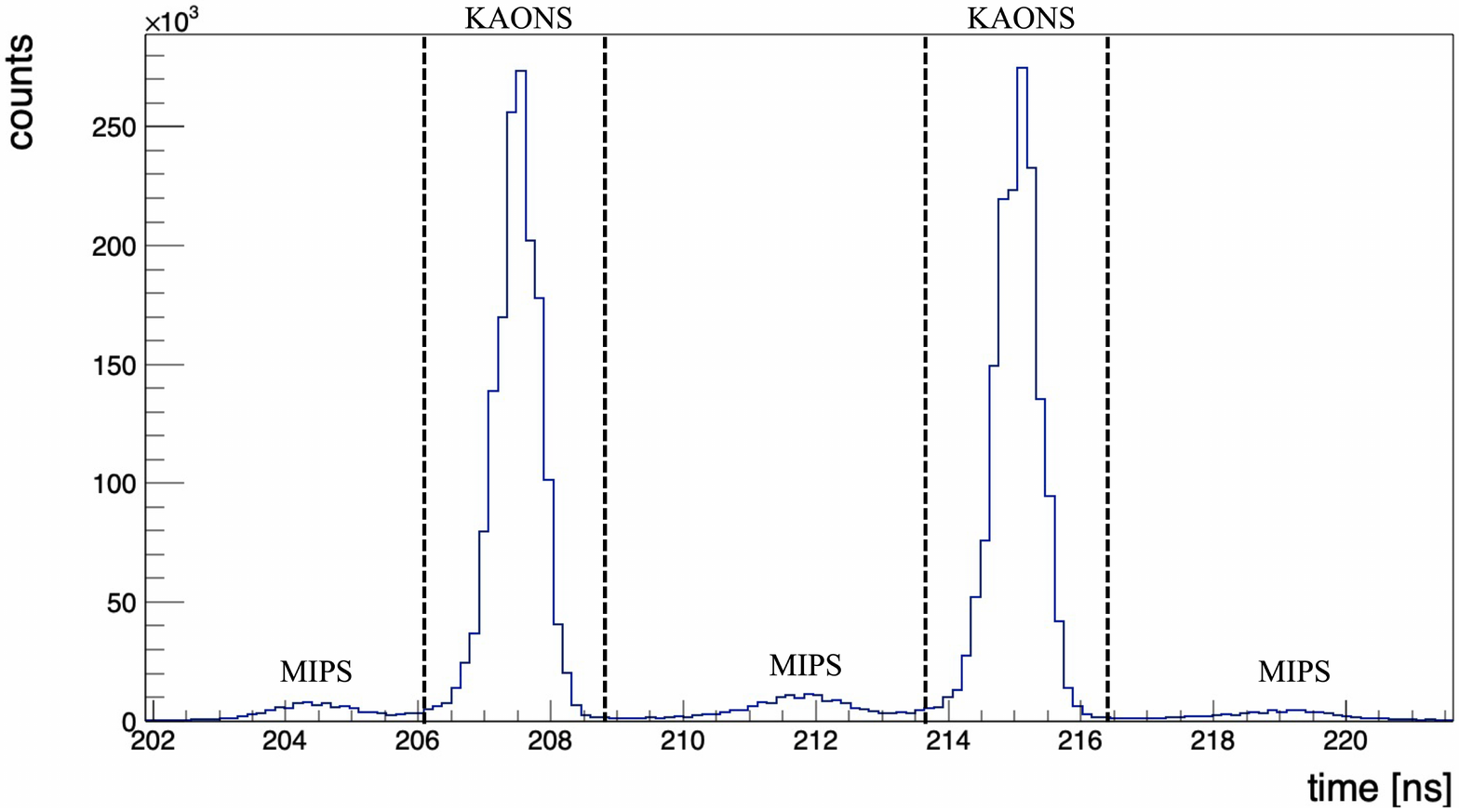}
\caption{\small{Time difference between a signal on the KT (up-down coincidence) and an SDD X-ray hit and the $\mathrm{1500\, ns}$ to $\mathrm{3450\, ns}$ acceptance window (top). Two-dimensional plot of the TDC for the upper and lower scintillators (mid) and its projection on the diagonal (bottom). The dotted lines represent the limits of the kaon selection time windows.}}
\label{kaon_mips}
\end{figure}

\noindent The background produced by the DA$\Phi$NE collider has two main components. The first one, mainly due to particles lost from the two rings (MIPs), is asynchronous with the $e^+e^-$ collisions and can be highly suppressed using the Kaon Trigger; the second one, the synchronous background, corresponding to electromagnetic showers and heavier particles correlated in time with the collisions and the subsequent hadronic and electromagnetic processes, and is more difficult to suppress. Additional veto systems are foreseen for this task in the final SIDDHARTA-2 setup.\\
The reduction of the asynchronous background is performed by selecting only the events occurring in time coincidence, within a $\mathrm{5\,\mu s}$ window, with a signal on the KT.
The time difference between the $K^-K^+$ coincidence in the KT (up-down coincidence) and the time of the X-ray detection by SDDs is shown in the top part of Fig. \ref{kaon_mips}, where the peak corresponds to X-ray signals on the SDDs in coincidence with the KT, while the flat distribution is the result of the uncorrelated events. The width of the peak is 950 ns (FWHM) and it reflects the drift time distribution of the electrons in the bulk of the SDDs. The region from $\mathrm{1500\,ns}$ to $\mathrm{3450\,ns}$, delimited by the red dashed lines, is the drift time window selection. This trigger selection suppresses the background by a factor $\simeq\,10^5$.\\ 
In addition to this selection, the time difference between the signals on each scintillator and the DA$\Phi$NE radiofrequency (RF) can be used to discriminate, via time of flight, whether an event on the KT is related to a kaon or to a MIP entering the target cell. \\
The middle pad of Fig. \ref{kaon_mips} shows the two-dimensional plot of the TDC for the upper and lower scintillators, while in the lower pad the projection on the diagonal is shown. The double KAONS/MIPs structure is due to the usage of the RF/2 signal as a time reference.
From this figure, one can see the kaon-related events, corresponding to the two main peaks, can be discriminated from the MIP-related ones. Among all triggered events only the kaon identified events, lying in the time windows within the dotted lines in Fig. \ref{kaon_mips}, are finally accepted. 

\begin{figure}[htbp]
\centering
\includegraphics[width=14 cm]{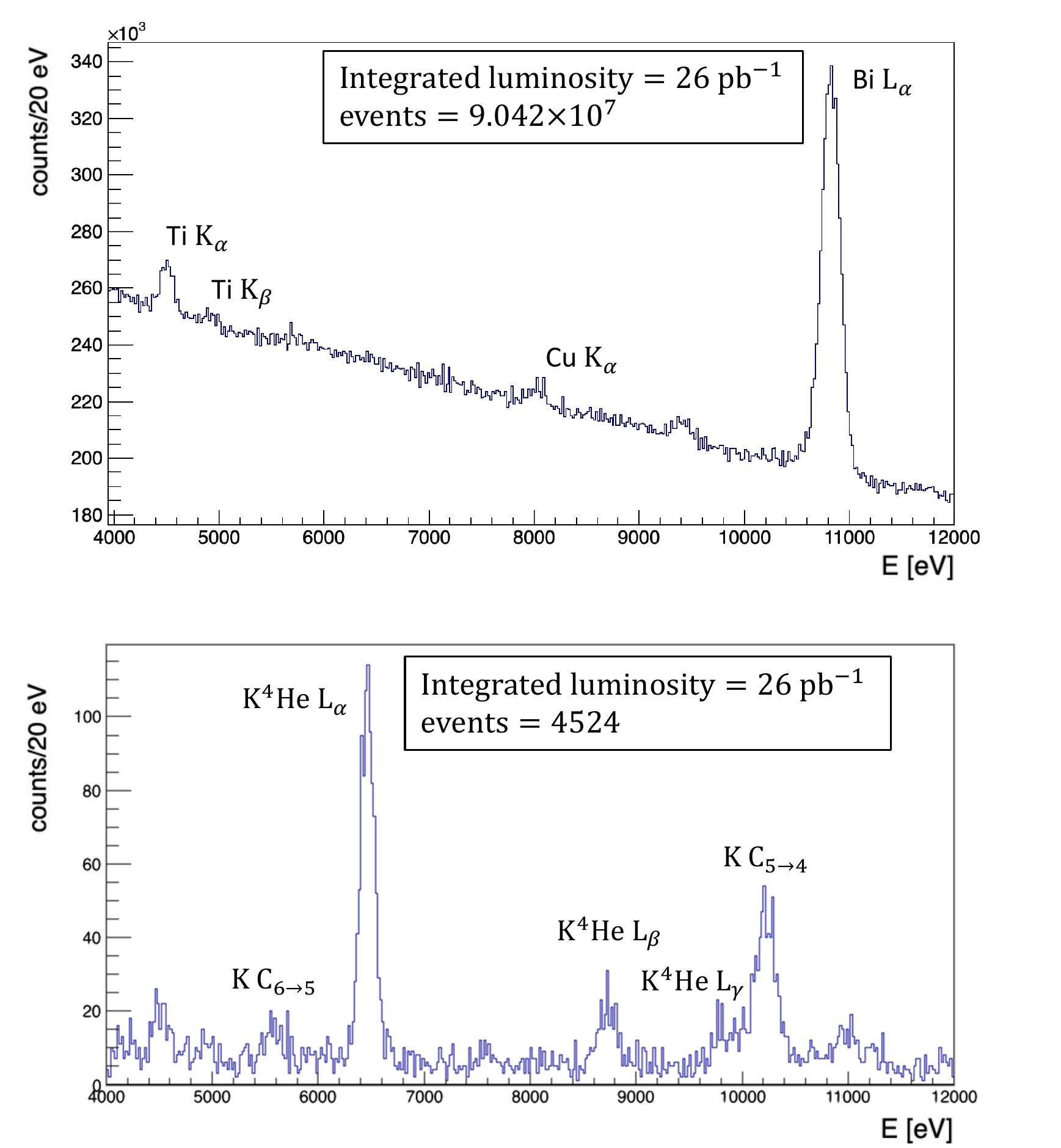}
\caption{\small{Spectra without (top) and with (bottom) KT selections, from which the $\simeq\,10^5$ rejection factor can be obtained (bottom). See text for details.}}
\label{drift}
\end{figure}

\noindent In the upper panel of Fig. \ref{drift}, the inclusive kaonic helium-4 energy spectrum corresponding to 26 $\mathrm{\,pb^{-1}}$ integrated luminosity, with no background rejection, is shown. The continuous contribution below the peaks is mainly due to the asynchronous component of the machine background, while the Ti and Cu peaks originate from the calibration foils installed inside the target and activated by particles lost from the beams. The Bi peak was instead produced by the activation of the alumina carrier behind the SDDs silicon wafer. The region of interest for the $K^4He(3d\rightarrow2p)$ transition measurement lies around 6.4 keV. In this region no peaks potentially affecting the result of the measurement are present.\\
In the bottom panel of Fig. \ref{drift}, the spectrum of the selected events is presented; here, the rejection capabilities of the KT can be appreciated; requiring the above mentioned triple coincidence and separating kaons and MIPs, a $\simeq\,10^5$ rejection factor is achieved and the $K^4He(4d,3d\rightarrow2p)$ peaks are clearly observed. 

\subsection{Degrader optimization}\label{deg_section}

\begin{figure}[htbp]
\centering
\mbox{\includegraphics[width=11.5 cm]{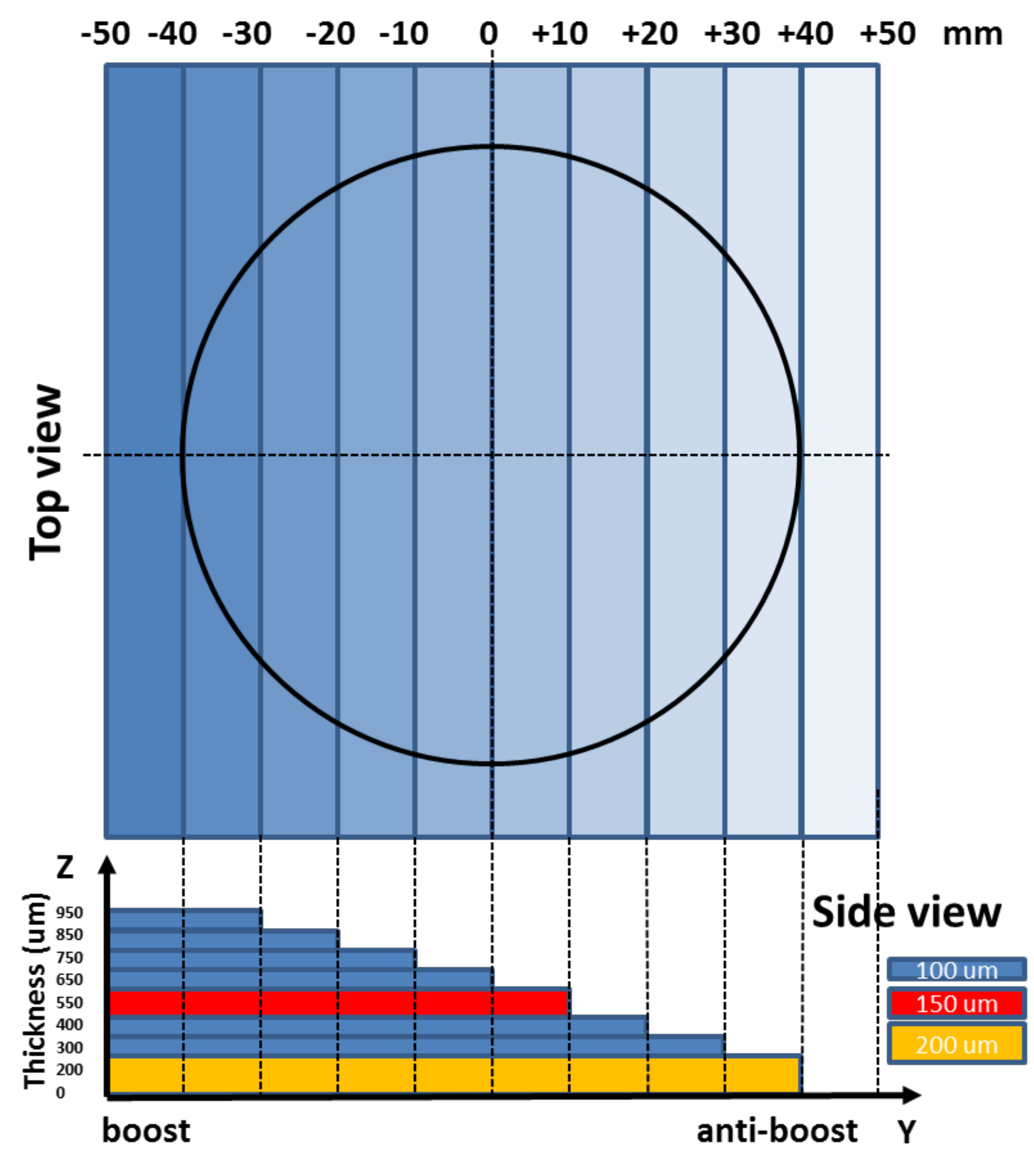}}
\caption{\small{The final Mylar degrader configuration: the circle represents the
size of the entrance window of the vacuum chamber; direction “Y” points to the outer side of
the DA$\Phi$NE ring, corresponding to the anti-boost side for kaons. The degrader has
eight steps to compensate for the boost effect, with thicknesses shown in the lower part of the figure.}}
\label{degrader}
\end{figure}

\begin{figure}[htbp]
\centering
\mbox{\includegraphics[width=15 cm]{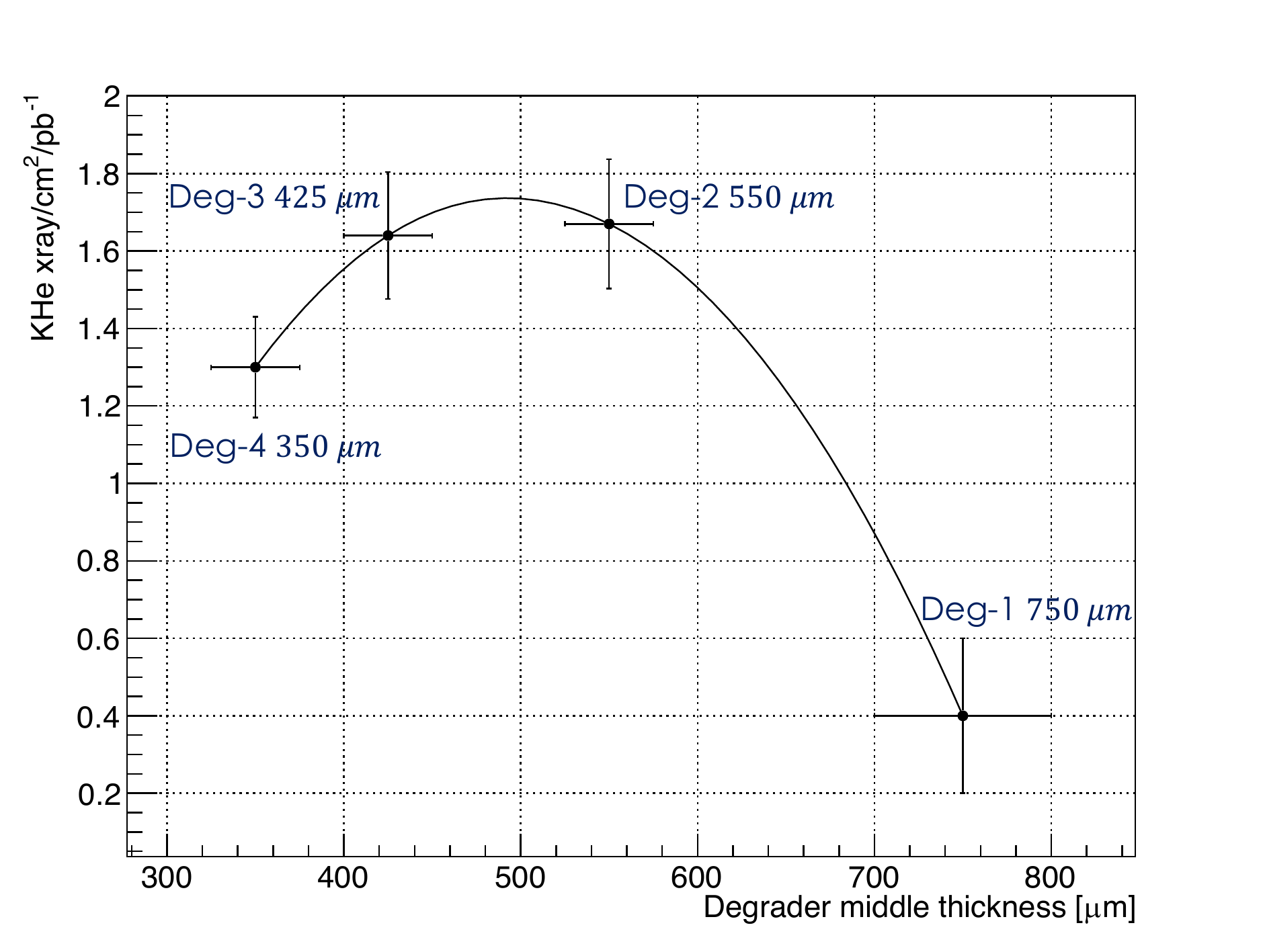}}
\caption{\small{Degrader optimization curve: the horizontal axis is the central thickness and the vertical one the corresponding $K^4He(3d\rightarrow2p)$ signal normalized by integrated luminosity and effective detection surface.}}
\label{deg_curve}
\end{figure}

Before entering the helium target, the kaons cross the beam pipe, the scintillators of the KT, the vacuum chamber, and target entrance windows; after passing through all these materials, the kaons need to be additionally slowed down to maximize the stopping fraction in the target.
For this purpose, a degrader is placed below the upper scintillator of the KT.\\
Since the trajectories of DA$\Phi$NE electron and positron beams cross, in the interaction point, at an angle of 50 mrad between each other, the center of mass of the subsequent $K^+K^-$ system receives a boost towards the center of the collider, which is reflected in the momentum distribution of the kaons.\\ 
To compensate for this effect and to obtain a uniform stopping  distribution of the kaons inside the target, a stepwise degrader is used. The degrader, which is schematically shown for the optimal configuration in Fig. \ref{degrader}, consists of 8 Mylar strips of $\mathrm{1\times9\,\, cm^2}$ each varying from $\mathrm{100\, \mu m}$ to $\mathrm{200\, \mu m}$ thickness ($\mathrm{550\, \mu m}$ in the central position, namely Y=0 in Fig. \ref{degrader}). The overall thickness of the eight-step degrader ranges from $\mathrm{200\,\mu m}$ to $\mathrm{950\, \mu m}$.\\
To optimize the shape and thickness of the degrader, MC simulations have been performed, as well as an experimental fine tuning, based on the amplitudes of the observed kaonic helium signals. For different degrader configurations, the number of $K^4He(3d\rightarrow2p)$ X-ray events, normalized to the integrated luminosity and effective detection surface was recorded and its trend as a function of the degrader central thickness is shown in Fig. \ref{deg_curve}.

\noindent The optimal degrader configuration, shown in Fig. \ref{degrader}, is extrapolated from the scan of Fig. \ref{deg_curve}. 
This optimization is an important and delicate operation since, as clearly shown by the curve, even a small variation of about 200 $\mathrm{\mu m}$ can drastically reduce the kaonic atoms signal.

\section{The new $K^4He(3d\rightarrow2p)$ shift and width measurement by SIDDHARTINO}\label{he_section}

\begin{figure}[htbp]
\centering
\mbox{\includegraphics[width=14 cm]{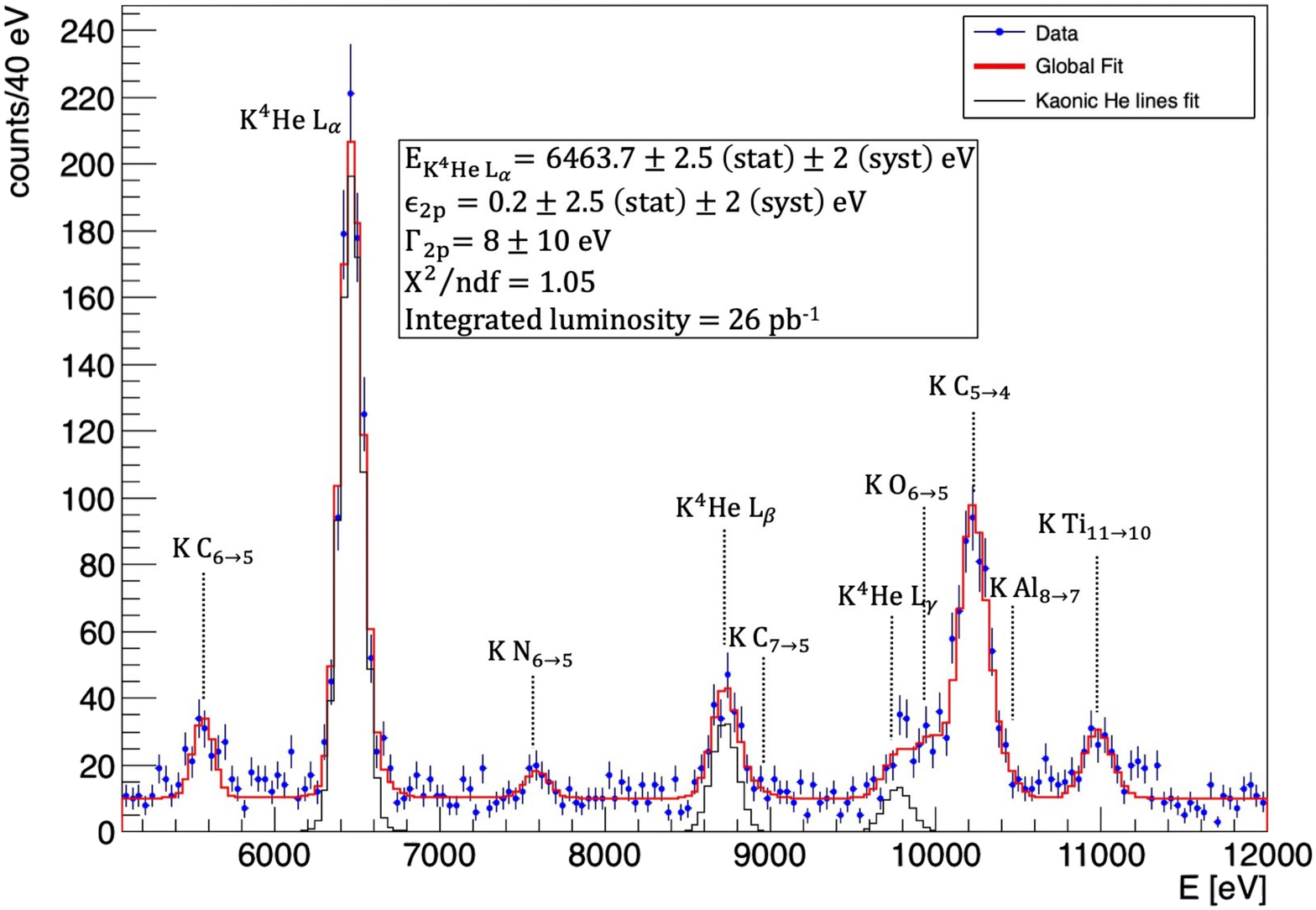}}
\caption{\small{Fit (red line) of the $K^4He$ energy spectrum. The L$\alpha$ peak is seen together with the L$\beta$ and L$\gamma$ ones (black lines). The peaks labeled as KN, KC, KAl, KTi (dotted lines) are the kaonic atoms lines produced by the kaons stopped in the Kapton ($C_{22}H_{10}O_{5}N_{2}$) walls of the target cell and in other parts of the setup (see text for details).}}
\label{He_spec}
\end{figure}

\noindent Fig. \ref{He_spec} shows the final $K^4He$ spectrum for the total integrated luminosity of 26 $\mathrm{\, pb^{-1}}$. The $K^4He$ L$\alpha$ line (3d$\rightarrow$2p) is visible together with the L$\beta$ (4d$\rightarrow$2p) and L$\gamma$ (5d$\rightarrow$2p) ones. Other lines, corresponding to kaonic carbon, nitrogen, and oxygen high-n transitions generated by kaons stopped in the target window made of Kapton ($C_{22}H_{10}O_{5}N_{2}$), were also detected, as well as lines from kaonic titanium and aluminum due to kaons stopping in the other components of the experimental apparatus.

\noindent The $K^4He$ peaks were fitted with a Voigt function \cite{SIDDHARTA:2012rsv}, the Gaussian part of which reproduces the detector energy response function, while the Lorentzian one describes the intrinsic linewidth of the transition. 
From the calibration procedure of the SDDs, it was found that the exponential low energy tails account for less than $1\,\%$ of the peak amplitude, and hence are negligible for the kaonic helium peaks \cite{2021tail}.  \\
The other peaks were fitted using only the Gaussian part since the shift and broadening due to the strong interaction are known to be negligible for $n>2$ levels \cite{Okada:2007ky,Friedman:1994hx}. The difference between the purely electromagnetic calculated transition energies and the experimental ones, representing the strong interaction induced shift, can be then set as a common parameter to all the $K^4He$ peaks.
Similarly to the shift parameter, also the strong interaction induced broadening of the 2p level, representing the Lorentzian contribution to the Voigt function, can be set as a common parameter to all the $K^4He$ lines. From the results of the fit we measured the strong interaction induced shift and width of the 2p level to be:
\begin{eqnarray}
&\epsilon_{2p}= \mathrm{E_{exp} - E_{e.m}} = \mathrm{0.2\pm2.5 (stat)\pm2 (syst)\ eV}& \\
&\Gamma_{2p} = \mathrm{8\pm10\ eV \,(stat)}&
\end{eqnarray}
The systematic errors on the shift value were evaluated from the linearity and the stability of the energy response of the SDDs; other possible contributions (e.g. kaon timing window selection, different background contributions to the fit function) are negligible. The systematic errors on the width are negligible (less than $\mathrm{0.1\,eV}$).\\
These results represent the most precise measurement of gaseous $K^4He$ (3d$\rightarrow$2p) transition  and confirm the experimental observations performed by the E570 \cite{Okada:2007ky} and the SIDDHARTA \cite{SIDDHARTA:2012rsv} experiments. 

\section{Conclusions}

The SIDDHARTINO experiment at the DA$\Phi$NE collider of INFN-LNF, the pilot run for the SIDDHARTA-2 experiment which aims to perform the first measurement ever of kaonic deuterium transitions to the fundamental level, has successfully been concluded during the commissioning phase of the collider in 2021.
During this run, a series of optimizations of the experimental apparatus, including that of the degrader, necessary for the optimization of the fraction of negatively charged kaons inside the cryogenic gaseous target, have been performed. The thin plastic degrader, shaped such as to take into account the boost of the $\Phi$-particles generating kaons with an energy-angle dependence, has been optimized at a precision level better than 100 $\mathrm{\mu m}$. The procedure exploited the kaonic helium-4 transitions to the 2p level measurements at an equivalent 1.5\% liquid helium density. Also, a set of data at half of the previous density were collected. \\ 
The analysis of the overall set of kaonic helium data, for an integrated luminosity of 26 $\mathrm{pb^{-1}}$, has moreover delivered the most precise measurement of kaonic helium-4 X-ray transitions in gas, in terms of shift and width of the 2p level induced by the strong interaction.
The SIDDHARTINO results put an even more stringent limit than the previous SIDDHARTA outcome \cite{SIDDHARTA:2012rsv}, and  exclude large shifts and widths. Such results, on one side, contribute to a better understanding of the strong interaction at low-energy in systems with strangeness, even if more precise results are mandatory. On the other side, the successful completion of the SIDDHARTINO run proved the innovative technologies and methodologies employed by the collaboration in studies of exotic atoms in a collider environment are sound and solid. It sets the ground for the kaonic deuterium measurement with the SIDDHARTA-2 experiment, which is presently being installed on the DA$\Phi$NE collider, aiming to start the data acquisition campaign in 2022 for an overall integrated luminosity of 800 $\mathrm{pb^{-1}}$. This experiment should provide a kaonic deuterium measurement of the same level of precision as the kaonic hydrogen one performed by SIDDHARTA \cite{Bazzi:2011}. \\
Other types of kaonic atom measurements with various radiation detectors are presently under consideration, to be proposed and performed after the SIDDHARTA-2 run. They could further contribute to a deeper understanding of the strong interaction in the low-energy regime in the strangeness sector, impacting particle and nuclear physics and also astrophysics \cite{Curceanu:2021oqj}.

\section{Acknowledgments}

We thank C. Capoccia from LNF-INFN and H. Schneider, L. Stohwasser, and D. Pristauz-Telsnigg from Stefan-Meyer-Institut for their fundamental contribution in designing and building the SIDDHARTA-2 setup.~We thank as well the DA$\Phi$NE staff for the excellent working conditions and permanent support. Part of this work was supported by the Austrian Science Fund (FWF): [P24756-N20 and P33037-N]; the Croatian Science Foundation under the project IP-2018-01-8570; EU STRONG-2020 project (grant agreement No.~824093), the EU Horizon 2020 project under the MSCA G.A. 754496 and the Polish Ministry of Science and Higher Education grant No. 7150/E-338/M/2018 and the Foundational Questions Institute and Fetzer Franklin Fund, a donor-advised fund of Silicon Valley Community Foundation (Grant No. FQXi-RFP-CPW-2008)

\section*{References}
\bibliography{iopart-num}

\end{document}